\documentstyle[12pt]{article}
\input{psfig}
\long\def\comment#1{}
\begin{document}
\title{Greek and Indian Cosmology: Review of Early History}

\author{Subhash Kak\thanks{
Louisiana State University,
Baton Rouge, LA 70803-5901, USA,
Email: {\tt kak@ece.lsu.edu}}}

\maketitle

\section{Introduction}

Greek and Indian traditions have profoundly
influenced modern science.
Geometry, physics, and
biology of the Greeks; 
arithmetic, algebra, and grammar of the Indians;
and 
astronomy, philosophy, and medicine of both
have played a key role in the creation of knowledge.
The interaction between the Indians and the Greeks after the
time of Alexander is well documented, but can we trace this
interaction to periods much before Alexander's time
so as to untangle
the earliest connections
between the two, especially as it concerns scientific ideas?

Since science is only one kind of
cultural expression, our search must encompass other items
in the larger matrix of cultural forms so as to obtain a
context to study the relationships.
There are some intriguing parallels
between the two but there are also
important differences.
In the ancient
world there existed much interaction through trade 
and evidence for this interaction has
been traced back to the third millennium BC, therefore
there was sure to have been a flow of ideas in different directions.

The evidence of interaction comes from the trade routes
between India and the West that were active during the
Harappan era.
Exchange of goods was doubtlessly accompanied by an 
exchange of ideas. Furthermore, some communities migrated 
to places away from their homeland. For example, an Indian settlement
has been traced in Memphis in Egypt based on the Indic themes of its
art.$^1$
The Indic element had a significant presence in West Asia
during the second millennium BC and later.$^2$
Likewise, Greek historians accompanying Alexander reported the
presence of Greek communities in Afghanistan.$^3$

In view of these facts,
the earlier view of the rise in
a vacuum
of Greek science cannot be maintained.$^4$
Indian science must have also benefited from outside
influences.

 The indebtedness
of the Greeks to the 
Babylonians and the Egyptians is now acknowledged, 
thanks to the decipherment of the Babylonian
tablets of the second millennium BC.
The first flowering of Greek philosophy took
place in Miletus, a trading centre of the Ionian Greeks on the Asiatic coast, 
where Greek and Asiatic cultures mingled. In addition to
the Babylonians 
antecedents of Greek science there existed
Indian antecedents.

In the desire amongst Eurocentric historians
to trace science and philosophy to Greece alone so that
somehow it would then appear as something uniquely ``Western,'' 
the question of the Asiatic basis to Greek science and philosophy 
the Asian prehistory of Western science is ignored. 
More recently there is grudging acknowledgement that the Babylonians
and the Egyptians
may have contributed to the ideas.$^5$

It is instructive to begin with the
similarities in the Greek and Indian sciences of the late
first millennium BC. Specifically, I shall consider the
similarities in geometry, astronomy, and medicine.

In geometry, it is striking that the same type of constructions are
used in the \'{S}ulbas\={u}tras and by Euclid
to prove the Theorem of Pythagoras.$^6$
In astronomy, the size assumed for the solar system is very
similar and the planetary orbits are explained based on
retrograde motions.$^7$
In medicine, Plato speaks of three humours with a central
role to the idea of breath ({\it pneuma}), when 
a similar three-do\'{s}a system around  breath ({\it pr\={a}\d{n}a})
is already a feature of the much older Vedic
thought.$^8$

These commonalities have led to four different kind of theories:

\begin{enumerate}

\item Viewing the Indian evidence as being later than the
earliest occurrence in Greece. 
This notion was used by some 19th century European
scholars to date Indian texts.$^9$

\item Showing a common earlier origin for the two sciences outside
of India and Greece, perhaps as the common heritage of 
the Indo-Europeans.
More recently, this common source has been seen in Central Asia.$^{10}$

\item Showing that evidence exists for the priority of Indian 
sciences over the corresponding Greek sciences.$^{11}$

\item
A theory of essential
independent origin for the sciences in India and Greece although
some general ideas may have been carried through trade caravans
from one region to another.$^{12}$

\end{enumerate}

I shall now take each one of these views separately.
Before I proceed I would like to emphasize that the objective
of this essay is not to show the priority of Indian science.
In fact, my own position favours the fourth theory in the list above.

Scientific
developments occur as a consequence of certain social and
material conditions.
It is because of India's early urbanization that
many Indian
scientific innovations occurred at a time 
that predates the early Greek scientific age.
But it does not follow that the Indian innovations were
directly linked to the corresponding innovations in 
Greece.
 
I shall begin by providing the essentials of the Indic
world-view which highlights its 
points of difference with the Greek world-view.
This is important to define the context in which a 
relationships between the two may be examined.
I shall, in particular, make the comparisons both in 
regard to abstract theory as in mathematics and astronomy and
in experiment and observation as in medicine.
My selection of topics is not exhaustive since my
objective is to point to issues of difference (as in
astronomy) and that of similarity (medicine).

\section{Indian science and its cosmology}

Indian archaeology and literature provides us with considerable layered
evidence related to the development of science.
The chronological time frame for this history is provided by the
archaeological record which has been traced in an unbroken tradition
to about 8000 BC. Prior to this we have records of rock paintings
that are believed to be considerably older. The earliest textual
source is the Rig Veda which is a compilation of very early material.
There are astronomical references in this and the
other Vedic books which recall events in the third
or the fourth millennium BC and earlier.  The discovery that
Sarasvati, the preeminent river of the Rig Vedic times, went dry around
1900 BC due to tectonic upheavels supports the view 
that the Rig Veda hymns recall events 
dated prior to this epoch. According to traditional history, 
Rig Veda is prior to 3100 BC.
The astronomical evidence related to winter solstices shows a
layered chronology of early Indian texts from the third to the
first millennia BC.$^{13}$

Indian writing (the so-called Indus script) goes back to the beginning of 
the third millennium BC but it has not yet been deciphered. However, 
statistical analysis shows that 
Br\={a}hm\={\i} (of which earliest records have  been traced
to 550 BC in Sri Lanka) evolved out of this writing.
The invention of the symbol for zero appears to have 
been made around 50 BC to 50 AD.$^{14}$

\subsubsection*{Vedic cosmology}

Briefly, the Vedic texts present a tripartite and recursive world view.  The
universe is viewed as three regions of earth, space, and sky which
in the human being are mirrored in the physical body, the breath 
(pr\={a}\d{n}a), 
and mind. The processes in the sky, on 
earth, and within the mind are taken to be connected.  The universe is 
also connected to the human mind, leading to the idea that introspection
can yield knowledge. 
The universe goes through
cycles of life and death.

The Vedic seers were aware that
all descriptions of the universe lead to logical paradox.  The one
category transcending all oppositions is {\it Brahman}.
Understanding the nature of consciousness was of paramount importance
in this view but this did not mean that other sciences were ignored.
Vedic ritual was a symbolic retelling of this conception.
The notable features of this world view are:

\begin{description}

\item
{\it 1. An Extremely Old and Large Cyclic Universe: }
The Vedas speak of an infinite universe, and the
Br\={a}hma\d{n}as (e.g. Pa\~{n}cavi\d{m}\'{s}a) mention very large yugas.
The recursive Vedic world-view requires that the universe itself go
through cycles of creation and destruction.  This view became a part of
the astronomical framework and ultimately very long cycles of billions
of years were assumed.  
The Pur\={a}\d{n}as speak of the universe going through cycles of creation and
destruction of 8.64 billion years, although there are
longer cycles as well.

\item
{\it 2. An Atomic World and the Subject/Object Dichotomy: }
According to the atomic doctrine of Ka\d{n}\={a}da, there
are nine classes of substances: ether, space, and time
that are continuous; four elementary substances (or particles)
called earth, air, water, and fire that are atomic;
and two kinds of mind, one omnipresent and another which
is the individual.
As in the systems of S\={a}\d{m}khya and Ved\={a}nta,
 a subject/object 
dichotomy is postulated.
The conscious subject is separate 
from the material reality but he is, nevertheless, able 
to direct its evolution. The atomic doctrine of Ka\d{n}\={a}da is much
more interesting than that of Democritus.

\item
{\it 3. Relativity of Time and Space: }
That space and time need not flow at the same rate for
different observers is
encountered in the
Br\={a}hma\d{n}a and Pur\={a}\d{n}a stories and in the Yoga
V\={a}si\d{s}\d{t}ha. Obviously, we are not speaking here of the
mathematical theory of relativity regarding an upper
limit to the speed of light, yet the consideration of 
time acting different to different observers is quite 
remarkable. 

Here's a passage on anomalous flow of time from the
Bh\={a}gavata Pur\={a}\d{n}a:$^{15}$ ``Taking his own daughter, Revati, Kakudmi 
went to Brahm\={a} in Brahmaloka, and inquired about a 
husband for her. When Kakudmi arrived there, Brahm\={a} 
was engaged in hearing musical performances by the
Gandharvas and had not a moment to talk with him. 
Therefore Kakudmi waited, and at the end of the performance 
he saluted Brahm\={a} and made his desire known. After hearing 
his words, Brahm\={a} laughed loudly and said to Kakudmi, 
`O King, all those whom you may have decided within the 
core of your heart to accept as your son-in-law have
passed away in the course of time. Twenty-seven caturyugas 
have already passed Those upon whom you may have decided 
are now gone, and so are their sons, grandsons. and other 
descendants. You cannot even hear about their names.'"

\item
{\it 4. Evolution of Life:}
The Mah\={a}bh\={a}rata (pre-400 BC) and the 
Pur\={a}\d{n}as have a chapter on creation and the rise of
mankind. It is said that man arose at the end of a
chain where the beginning was with plants and various
kind of animals. Here's the quote from the Yoga
V\={a}si\d{s}\d{t}ha:$^{16}$ "I remember that once upon a time there was 
nothing on this earth, neither trees and plants, nor 
even mountains.  For a period of eleven thousand years 
(four million earth years) the earth was covered by lava. 
Then demons (asuras) ruled the earth; they were 
deluded and powerful. The earth was their playground.
And then for a very long time the whole earth was 
covered with forests, except the polar 
region. Then there arose great mountains, but without
any human inhabitants. For a period of ten thousand 
years (4 million earth years) the earth was covered 
with the corpses of the asuras."

Vedic evolution is not like Darwinian
evolution; it has a different focus.  The urge to 
evolve into higher forms is taken to be inherent in 
nature.  A system of an evolution from inanimate to 
progressively higher life is taken to be a consequence
of the different proportions of the three basic attributes
of sattva, rajas, and tamas.

The doctrine of the three constituent qualities
plays a very important role in
the S\={a}\d{m}khya physics and metaphysics.
In its undeveloped state, cosmic matter has these qualities
in equilibrium.
As the world evolves, one or the other of these become
preponderant in different objects or beings,
giving specific character to each.

\item
{\it 5. A Science of Mind, Yoga: }
Inner science, described in the Vedic books and systematized 
by Pata\~{n}jali in his Yoga-s\={u}tras is a very sophisticated 
description of the nature of the human mind and its capacity. 
It makes a distinction between memory, states of awareness, 
and the fundamental entity of consciousness.  It puts the 
analytical searchlight on mind processes, and it does so with 
such clarity and originality that it continues to influence 
people all over the world. 

Several kinds of yoga are described. They provide a means
of mastering the body-mind connection. Indian music and
dance also has an underlying yogic basis.

\item
{\it 6. Binary Numbers, Infinity: }
A binary number system was used by Pi\.{n}gala$^{17}$ (450 BC, if 
we accept the tradition that he was P\={a}\d{n}ini's brother) 
to represent Vedic metres. The structure of this 
number system may have helped in the invention of the sign for
zero.
Without this sign, mathematics
would have languished. It is of course true that the binary
number system was independently invented by Leibnitz in 1678,
but the fact that the rediscovery had to wait almost 
2,000 years only emphasizes the originality of Pi\.{n}gala's idea.

The idea of infinity is found in the 
Vedas itself. It was correctly understood as one where addition
and subtraction of infinity from it leaves it unchanged.

\item
{\it 7. A Complete Grammar, Limitation of Language:}
The A\d{s}\d{t}\={a}dhy\={a}y\={\i}, the grammar of the Sanskrit 
language  by P\={a}\d{n}ini (450 BC), 
describes the entire language in 4,000 algebraic rules. 
The structure of this grammar contains a meta-language, 
meta-rules, and other technical devices that make this 
system effectively equivalent to the most powerful 
computing machine.$^{18}$ No grammar of similar power has 
yet been constructed for any other language. The 
famous American scholar Leonard Bloomfield called
P\={a}\d{n}ini's achievement as "one of the greatest monuments of
human intelligence."

The other side to the discovery of this grammar is the
idea that language (as a formal system) cannot describe
reality completely. This limitation of language
is why reality can only be experienced and never
described fully.

\end{description}

Many aspects of the Indian scientific system have
parallels in modern science. Indian ideas have found
special resonance amongst theoretical physicists 
and psychologists. 

Knowledge was classified in two ways: the lower or dual; 
and the higher or unified.  The seemingly irreconciliable
worlds of the material and the conscious were taken as
aspects of the same transcendental reality.

The idea of complementarity was at the basis of the systematization of
Indian philosophic traditions as well, so that complementary approaches
were paired together.  This led to the groups of: logic (Ny\={a}ya) and physics
(Vai\'{s}e\d{s}ika), cosmology (S\={a}\d{m}khya) and psychology (Yoga), 
and language
(M\={\i}ma\d{m}s\={a}) and reality (Ved\={a}nta).  Although these philosophical schools
were formalized in the post-Vedic age, we find the basis of these ideas
in the Vedic texts.

The S\={a}\d{m}khya and the Yoga systems take the mind as consisting of five
components:  manas, aha\d{m}k\={a}ra, citta, buddhi, and \={a}tman.  
Manas is the lower mind which collects sense impressions.  Aha\d{m}k\={a}ra is
the sense of I-ness that associates some perceptions to a subjective
and personal experience.  Once sensory impressions have been related to
I-ness by aha\d{m}k\={a}ra, their evaluation and resulting decisions are
arrived at by buddhi, the intellect.  
Citta is the memory bank of the mind.  These memories constitute the
foundation on which the rest of the mind operates.  But citta is not
merely a passive instrument.  The organization of the new impressions
throws up instinctual or primitive urges which creates different
emotional states.
This mental complex surrounding the innermost aspect of consciousness is
the \={a}tman, the self, or brahman.  

\subsubsection*{Logic}

The objective of the Ny\={a}ya is anv\={\i}k\d{s}ik\={\i}, or critical inquiry. The 
beginnings of it go into
the Vedic period, but its first systematic elucidation is due to Gautama 
in  his Ny\={a}ya
S\={u}tra dated to 3rd century BC. The
text begins with the nature of doubt and the means of proof. Next it 
considers self, body, senses and their objects, cognition and mind. It describes 
the cognizing
human in terms of volition, sorrow, suffering and liberation. 
The most important early commentary on this text is the 
Ny\={a}ya
Bh\={a}\d{s}ya of V\={a}tsy\={a}yana.

The Ny\={a}ya is also called pram\={a}\d{n}a \'{s}\={a}stra 
or the science of correct knowledge. Knowing
is based on four conditions: 1) The subject or the pram\={a}t\d{r}; 2) 
The object or the prameya
to which the process of cognition is directed; 3) The cognition or the 
pramiti; and 4) the
nature of knowledge, or the pram\={a}\d{n}a.

The Ny\={a}ya system supposes that we are so constituted so as to seek truth. 
Our minds are
not empty slates; the very constitution of our mind provides some knowledge 
of the
nature of the world. The four pram\={a}\d{n}as through which correct knowledge is 
acquired are:
pratyak\d{s}a or direct perception, anum\={a}na or inference, upam\={a}na 
or analogy, 
and \'{s}abda
or verbal testimony.

The function of definition in the Ny\={a}ya is to state essential nature 
(svar\={u}pa) that
distinguishes the object from others. Three fallacies of definition are 
described: ativy\={a}pti,
or the definition being too broad as in defining a cow as a horned animal; 
avy\={a}pti, or too
narrow; and asambhava, or impossible.

Gautama mentions that four factors are involved in direct perception: the 
senses
(indriyas) , their objects  (artha), the contact of the senses and the 
objects (sannikar\d{s}a),
and the cognition produced by this contact (j\~{n}\={a}na). 
The five sense organs, 
eye, ear, nose,
tongue, and skin have the five elements light, ether, earth, water, and air 
as their field,
with corresponding qualities of colour, sound, smell, taste and touch.

Manas or mind mediates between the self and the senses. When the manas is 
in contact
with one sense-organ, it cannot be so with another. It is therefore said to 
be atomic in
dimension. It is because of the nature of the mind that our experiences are 
essentially
linear although quick succession of impressions may give the appearance of 
simultaneity.

Objects have qualities which do not have existence of their own. The color 
and class
associated with an object are secondary to the substance. According to 
Gautama, direct
perception is inexpressible. Things are not perceived as bearing a name. 
The conception
of an  object on hearing a name is not direct perception but verbal cognition.

Dharmak\={\i}rti, a later Ny\={a}ya philosopher, recognizes four kinds of perception:
sense-perception, mental perception (manovij\~{n}\={a}na), 
self-consciousness, and yogic
perception. Self-consciousness is a perception of the self through its 
states of pleasure
and pain. In yogic perception, one is able to comprehend the universe in 
fullness and
harmony.

Not all perceptions are valid. Normal perception is subject to the 
existence of 1) the
object of perception, 2) the external medium such as light in the case of 
seeing, 3) the
sense-organ, 4) the mind, without which the sense-organs cannot come in 
conjunction with
their objects, and 5) the self. If any of these should function improperly, 
the perception
would be erroneous. The causes of illusion may be do\d{s}a (defect  in the 
sense-organ), sa\d{m}prayoga 
(presentation of only part of an object), or sa\d{m}sk\={a}ra (habit 
based on
irrelevant recollection).

Anum\={a}na (inference) is knowledge from the perceived about the unperceived. The
relation between the two may be of three kind: the element to be inferred 
may be the
cause or the effect of the element perceived, or the two may be the joint 
effects of
something else.

The Ny\={a}ya syllogism is expressed in five parts: 1) pratij\~{n}\={a}, or the 
proposition: the house
is on fire; 2) hetu, or the reason: the smoke; 3) ud\={a}hara\d{n}a, the example: 
fire is
accompanied by smoke, as in the kitchen; 4) upanaya, the application: as in 
kitchen so for
the house; 5) nigamana, the conclusion: therefore, the house is on fire. 
This recognizes
that the inference derives from the knowledge of the universal relation 
(vy\={a}pti) and its
application to the specific case (pak\d{s}adharmat\={a}). There can  be no 
inference unless
there is expectation (\={a}k\={a}\.{n}k\d{s}\={a}) about the 
hypothesis which is expressed in terms of the proposition.

The minor premise (pak\d{s}adharmat\={a}) is a consequence of perception, whereas 
the major
premise (vy\={a}pti) results from induction. But the universal proposition 
cannot be arrived
at by reasoning alone. Frequency of the observation increases the 
probability of the
universal, but does not make it certain. 

The Ny\={a}ya system lays stress on antecedence in its view of causality. But 
both cause and
effect are viewed as passing events. Cause has no meaning apart from 
change; when
analyzed, it leads to a chain that continues without end. Causality is 
useful within the
limits of experience, but it cannot be regarded as of absolute validity. 
Causality is only a
form of experience.
The advancement of knowledge is from upam\={a}na, or comparison, with 
something else
already well known. The leads us back to induction through alaukika 
pratyak\d{s}a at the
basis of the understanding.

\'{S}abda, or verbal testimony, is a chief source of knowledge. The meaning of 
words is by
convention. The word might mean an individual, a form, or a type, or all 
three. A
sentence, as a collection of words, is cognized from the trace 
(sa\d{m}sk\={a}ra) 
left at the end of
the sentence.
Knowledge is divided into cognitions which are not reproductions of former 
states of
consciousness (anubhava) and those which are (sm\d{r}ti). Memory is said to 
arise from a
contact of the \={a}tman with the manas and the trace left by the previous 
experience. The
impression is the immediate cause of the recollection, whereas recognition 
of identity
requires an inductive leap.

The Ny\={a}ya speaks of errors and fallacies arising by interfering with the 
process of correct
reasoning. The Ny\={a}ya attacks the Buddhist idea that no knowledge is certain 
by pointing
out that this statement itself contradicts the claim by its certainty. 
Whether cognitions
apply to reality must be checked by determining if they lead to successful 
action. Pram\={a},
or valid knowledge, leads to successful action unlike erroneous knowledge 
(viparyaya).

The Ny\={a}ya accepts the metaphysics of the Vai\'{s}e\d{s}ika. 
It is asserted that 
the universe has
certain elements that are not corporeal. The subjective cognitions and 
feelings which are
part of the individual's consciousness are transitory and, therefore, they 
cannot be
associated with substances. They are viewed as qualities associated with 
the \={a}tman.

\subsubsection*{Physics and chemistry}

In the Vai\'{s}e\d{s}ika system 
atoms combine to form different kinds of molecules which 
break up under the influence of heat.  The molecules come to 
have different properties based on the influence of various 
potentials (tanm\={a}tras).

Heat and light rays are taken to consist of very
small particles of high velocity.
Being particles, their velocity is finite.
The gravitational force is perceived as a wind.
The other forces are likewise mediated by atoms of
one kind or the other.

Indian chemistry developed many different alkalis, acids
and metallic salts by processes of calcination and
distillation, often motivated by the need to
formulate medicines. Metallurgists developed efficient techniques
of extraction of metals from ore. 

\subsubsection*{Geometry and mathematics}

Indian geometry began very early in the Vedic period in altar
problems as in the one where the circular altar (earth) is to be made equal
in area to a square altar (heavens).  Two aspects of the ``Pythagoras'' 
theorem are described in the \'{S}ulbas\={u}tra texts by Baudhayana and others.  
The geometric problems are often presented with
their algebraic counterparts. The solution
to the planetary problems also led to the development of algebraic methods.

\subsubsection*{Astronomy}

Using hitherto neglected texts, an astronomy of the third millennium 
BC has been discovered recently.  Y\={a}j\~{n}avalkya (1800 BCE ?) knew of a 
95-year cycle to harmonize the motions of the sun and the moon and 
he also knew that the sun's circuit was asymmetric.

Astronomical numbers played a central role in Vedic ritual.$^{19}$
Part of the ritual was to devise geometrical schemes related 
to the lengths of the solar and the lunar years. 
The organization of the Vedic books was also according to an
astronomical code.  To give just one example, the total number
of verses in all the Vedas is 20,358 which equals 261 $\times$ 78, a product
of the sky and atmosphere numbers of Vedic ritual!  

The second millennium text Ved\={a}\.{n}ga Jyoti\d{s}a of 
Lagadha went beyond the earlier
calendrical astronomy to develop a
 theory for the mean motions of the sun and the moon.  This marked the
beginnings of the application of mathematics to the motions of the
heavenly bodies.
An epicycle theory was used to explain planetary motions.
Later theories consider the motion of the planets with respect
to the sun, which in turn is seen to go around the earth.

The cosmological descriptions of the Pur\={a}\d{n}as are somewhat
confusing. Their objective is to describe the inner cosmos of the
individual in a manner that parallels the outer cosmos.
With the earth taken to be at the level of the navel with 
concentric continents around the central axis (spine), the
sun and the moon are at the north in the head and the
planets are viewed somewhat beyond. In this conception of
the inner cosmos, the earth is taken to be 500 million yojanas
(to be equated to the size of the human body) and the sun,
as a center within the brain is a tiny 9,000 yojanas in length.
The earth (the human body) is taken to rest on a turtle
(the planet earth).

\subsubsection*{Medicine}

\={A}yurveda, the Vedic system of medicine, views health as harmony between 
body, mind and soul.  It deals not only with the body, but also with 
psychological and spiritual health. Its two most famous texts belong to the 
schools of Caraka and Su\'{s}ruta. According to Caraka, health and disease 
are not predetermined and life may be prolonged by human
effort. Su\'{s}ruta defines the purpose of medicine to cure the diseases of 
the sick, protect the healthy, and to prolong life.

The beginnings of medicine may be traced to the \d{R}gveda, since it speaks of 
the bhi\d{s}aj, or physician, in connection with setting a broken bone. From 
other references the bhi\d{s}aj or vaidya emerges as a healer of disease and 
expert in herbs. The twin gods \={A}\'{s}vins are particularly associated with 
healing of blindness, lameness, and leprosy. They give an artificial leg to 
a hero who has lost a leg in battle.  They are also associated with 
rejuvenation. Soma is another healing deity.
In many contexts, Indra, Agni, and Soma represent 
the three dh\={a}tus of air, fire and water.
The Garbha Upani\d{s}ad describes the body as consisting of five elements
(with further groups of five as in S\={a}\d{m}khya), supported on six (the
sweet, sour, salt, bitter, acid and harsh juices of food), endowed with
six qualities, made up of seven tissues, three do\d{s}as, and
twice-begotten (through father and mother). It further adds that the
head has four skull-bones, with sixteen sockets on each side. It says
that the body has 107 joints, 180 sutures, 900 sinews, 700 veins, 500
muscles, 360 bones, and 45 million hairs.

According to the Pra\'{s}na Upani\d{s}ad, the number of veins is
727,210,201.  There are 101 chief veins, each with 100 branch veins, to
each of which are 72,000 yet smaller tributary veins.
In Ch\={a}ndogya Upani\d{s}ad, organisms are divided into three classes based
on their origin: born alive (from a womb), born from an egg, and born
from a germ.

According to the tradition, there existed six schools of medicine,
founded by the disciples of the sage Punarvasu \={A}treya. Each of these 
disciples Agnive\'{s}a, Bhela, Jat\={u}kar\d{n}a, 
Par\={a}\'{s}ara, H\={a}r\={\i}ta, and K\d{s}\={a}rap\={a}\d{n}i 
composed a Sa\d{m}hit\={a}.  Of these, the one composed by Agnive\'{s}a
 was supposed 
to be the best. The Agnive\'{s}a
 Sa\d{m}hit\={a} was later revised by Caraka and it 
came to be known as Caraka Sa\d{m}hit\={a}. \={A}yurveda is traditionally divided into 
eight branches which, in Caraka's scheme are:  1.
S\={u}trasth\={a}na, general principles; 2. Nid\={a}nasth\={a}na, 
cause of disease; 3.
Vim\={a}nasth\={a}na, diagnostics; 
4. \'{S}ar\={\i}rasth\={a}na, anatomy and embryology;
5. Indriyasth\={a}na, prognosis; 6. Cikits\={a}sth\={a}na, therapeutics; 7.
Kalpasth\={a}na, pharmaceutics; and 8. Siddhisth\={a}na, successful
treatment.

In the Caraka school, the first teacher was Bharadv\={a}ja, whereas in
the Su\'{s}ruta school, the first person to expound \={A}yurvedic knowledge
was Dhanvantari in the form of the king Divod\={a}sa. The Caraka 
and Su\'{s}ruta Sa\d{m}hit\={a} are compendiums of two 
traditions rather than texts composed by single authors. The beginnings of 
these traditions must go to the second millennium BC if not earlier because 
of the parallel information obtained in the Vedic Sa\d{m}hit\={a} and the 
description in the Mah\={a}bh\={a}rata. There is much that is common in the two 
texts, except that the Su\'{s}ruta Sa\d{m}hit\={a} is richer in the field of 
surgery.  Part of the original Caraka Sa\d{m}hit\={a} is lost, and the current 
version has several chapters by the Kashmiri scholar D\d{r}\d{d}habala.

An attempt to reconcile the texts of Caraka and Su\'{s}ruta was made by
V\={a}gbha\d{t}a the Elder in second century BC in his A\d{s}t\={a}\.{n}ga 
Sa\.{n}graha.  The
works of Caraka, Su\'{s}ruta, and the Elder V\={a}gbha\d{t}a are considered
canonical and reverentially called the V\d{r}ddha Tray\={\i}, "the triad of
ancients." Later, V\={a}gbha\d{t}a the Younger wrote the
A\d{s}\d{t}\={a}\.{n}ga H\d{r}daya Sa\d{m}hit\={a} which is a lucid presentation of the \={A}yurveda
giving due place to the surgical techniques of Su\'{s}ruta. In the eighth
century, M\={a}dhava wrote his Nid\={a}na, which soon assumed a position of
authority. 

Health in \={A}yurveda is considered to be a balance of the three do\d{s}as or
primary forces of pr\={a}\d{n}a or v\={a}ta (air), agni or pitta (fire) and soma or
kapha (water and earth).  The five elements of the S\={a}\d{m}khya enumeration,
that is earth, water, fire, air, and ether, in different combinations
constitute the three body do\d{s}as: v\={a}ta (air and ether), pitta (fire)
and kapha (earth and water).

The trido\d{s}a or tridh\={a}tu theory of \={A}yurveda has sometimes been 
misunderstood to imply that v\={a}ta, pitta, and kapha literally 
mean air, bile, and phlegm, which are the ordinary 
physiological meanings of the terms. In reality,
v\={a}ta stands for the principle of motion, cell development
in general, and the functions of the central nervous system in
particular.  Pitta signifies the function of metabolism, including
digestion and formation of blood, various secretions and excretions
which are either the means or the end product of tissue combustion.
Kapha represents functions of cooling, preservation and heat
regulation. The imbalance of these elements leads to
illness. The predominance of one or the other
represents different psychological types.

Each of the do\d{s}as is recognized to be of five kinds. V\={a}ta appears as pr\={a}\d{n}a 
(governing respiration), ud\={a}na (for uttering sounds and speaking), sam\={a}na 
(for separating the digested juice), vy\={a}na (carrying fluids including blood 
to all parts of the body), and ap\={a}na (expelling waste products). Pitta 
appears as p\={a}caka (for digestion and imparting heat), ra\~{n}jaka 
(imparting 
redness to the chyle and blood), s\={a}dhaka (increasing the power of the 
brain), \={a}locaka (strengthen vision), and bhr\={a}jaka (improve complexion). 
Kapha has kledaka (moists food), avalambaka (imparts energy and strength), 
bodhaka (enables tasting), tarpaka (governs the eye and other sensory 
organs), and \'{s}le\d{s}maka (acts as lubrication).

Every substance (animal, vegetable or mineral) is a dravya with 
properties in different proportions: rasa, gu\d{n}a, 
v\={\i}rya, vip\={a}ka, and prabh\={a}va. The gu\d{n}as are qualities such as heat, cold, 
heaviness, lightness and so on in a total of twenty types. Of the twenty 
gu\d{n}as, 
 heat (u\d{s}\d{n}a) and cold (\'{s}\={\i}ta) 
are the most 
prominent. V\={\i}rya is generative energy that may also be hot or cold.

Vip\={a}ka may be understood as the biochemical transformations of food whereas 
prabh\={a}va is the subtle effect on the body of the substance. Food is 
converted into rasa by the digestive action of j\={a}\d{t}har\={a}gni, the fire in 
the stomach. Rasas are six in number: madhura, \={a}mla, lava\d{n}a, 
tikta, ka\d{t}u, 
and ka\d{s}\={a}ya. Each rasa is a result of the predominance of two elements and 
each is recognized by the taste. The knowledge of the rasas is 
important in therapeutics. Madhura, \={a}mla and lava\d{n}a work 
well against v\={a}ta; madhura, tikta, and ka\d{s}\={a}ya against pitta; and 
ka\d{t}u and 
ka\d{s}\={a}ya against kapha.

The five elements in various proportions are said to form seven kinds
of tissue (dh\={a}tu). These are: rasa (plasma), rakta (blood), m\={a}\d{m}sa
(flesh), medas (fat), asthi (bone), majj\={a} (marrow), and \'{s}ukra (semen).
The activity of the dh\={a}tu is represented by ojas (vitality) or bala
(strength). Ojas is mediated through an oily, while fluid that
permeates the whole body.  The functions of the vital organs like the
heart, brain, spleen, and liver relate to the flow
and exchange of tissues. The heart is the chief receptacle
of the three chief fluids of the body: rasa, rakta, and ojas.

The Bhela Sa\d{m}hit\={a}, ancient like the Caraka and Su\'{s}ruta Sa\d{m}hit\={a} but
available only in fragments, considers the brain to be the center of
the mind. It distinguishes between manas (mind) with its seat in the
brain and the citta, consciousness, with its seat in the heart.

\subsubsection*{Training a Vaidya}

The 
\={A}yurvedic vaidya must master eight branches: k\={a}y\={a}cikits\={a} 
(internal medicine),
\'{s}alyacikits\={a} (surgery including anatomy), \'{s}\={a}l\={a}kyacikits\={a} (eye, ear,
nose, and throat diseases), kaumarabh\d{r}tya (pediatrics), bh\={u}tavidy\={a}
(psychiatry, or demonology), and agada tantra (toxicology), ras\={a}yana 
(science of rejuvenation), and v\={a}jikara\d{n}a (the science of fertility).

In addition, the vaidya
was expected to know ten arts that were considered
indispensable in the preparation and application of medicines:
distillation, operative skills, cooking, horticulture, metallurgy,
sugar manufacture, pharmacy, analysis and separation of minerals,
compounding of metals, and preparation of alkalis.  The teaching of
anatomy, physiology, pathology, microbiology, and pharmacology was
done during the instruction of relevant clinical subjects.  For
example, teaching of anatomy was a part of the teaching of surgery,
embryology was a part of training in pediatrics and obstetrics, and the
knowledge of physiology and pathology was interwoven in the teaching of
all the clinical disciplines.

The initiation ceremony of the physician was called upanayana and it
involved the teacher leading the student three times around the sacred
fire.  This ceremony made the student thrice-born (trija),
distinguished from the twice-born (dvija) non-physicians.

At the closing of the initiation, the guru gave a solemn address to the
students where the guru directed the students to a life of chastity,
honesty, and vegetarianism. The student was to strive with all his
being for the health of the sick. He was not to betray patients for his
own advantage. He was to dress modestly and avoid strong drink. He was
to be collected and self-controlled, measured in speech at all times.
He was to constantly improve his knowledge and technical skill. In the
home of the patient he was to be courteous and modest, directing all
attention to the patient's welfare. He was not to divulge any knowledge
about the patient and his family. If the patient was incurable, he was
to keep this to himself if it was likely to harm the patient or
others.

Su\'{s}ruta speaks of a similar address to the initiates. In addition, the
student was told to wear an ochre robe. The teacher also took an oath:
``If you behave well and I fail to take care of you, that will be my sin
and my learning will be of no avail."

The normal length of the student's training appears to have been seven
years. Before graduation, the student was to pass a test. But the
physician was to continue to learn through texts, direct
observation (pratyak\d{s}a), and through inference (anum\={a}na). In addition,
the vaidyas attended meetings where knowledge was exchanged. The
doctors were also enjoined to gain knowledge of unusual remedies from
hillsmen, herdsmen, and forest-dwellers.

The qualified vaidya on his rounds from house to house was attended by
an assistant who carried his bag of instruments and herbs. He was clad
in white, shod in sandals, with a staff in hand, he had a servant
follow him with a parasol.  He would also see patients at his own house
where he also had a storeroom with drugs and instruments. He compounded
many of the drugs himself from herbs with the help of his assistant.

The vaidya was assisted by nurses (paric\={a}raka). Su\'{s}ruta lists the
following qualities of the nurse: devotion and friendliness,
watchfulness, not inclined to disgust, and knowledgeable to follow the
instructions of the doctor.

There existed governmental control of the medical profession.  Su\'{s}ruta
hints of this when he mentions that a quack kills people out of greed,
because of the fault of the king.
There is reference to free hospitals in ancient India.

\subsubsection*{Dissection and Surgery}

Su\'{s}ruta laid great emphasis on direct observation and learning through
dissection (avaghar\d{s}ana). In preparation of dissection the excrements
of the selected dead body were cleaned. The body was now covered with a
sheath of grass and left to decompose in the still waters of a pool.
After seven days, the student was instructed to scrape off the skin and
carefully observe the internal organs of the body.

Su\'{s}ruta classified surgical operations into eight categories:  incision
(chedya), excision (bhedya), scarification (lekhya), puncturing
(vedhya), probing (esya), extraction (\={a}h\={a}rya), evacuation and
drainage (vi\'{s}r\={a}vya), and suturing (s\={\i}vya). Su\'{s}ruta lists 101 blunt
and 20 sharp instruments that were used in surgery, instructing that
these should be made of steel and kept in a portable case with a
separate compartment for each instrument. Fourteen types of bandages
were described. Surgical operations on all parts of the body were
described. These include laparotomy, craniotomy, caesarian section,
plastic repair of the torn ear lobule, cheiloplasty, rhinoplasty,
excision of cataract, tonsillectomy, excision of laryngeal polyps,
excision of anal fistule, repair of hernias and prolapse of rectum,
lithotomy, amputation of bones, and many neurosurgical procduces.

Medications were used for pre-operative preparation and medicated oils
were used for dressing of wounds. Ice, caustics and cautery were used
for haemostasis. Medicated wines were used before and after surgery to
assuage pain.  A drug called sammohini was used to make the patient
unconscious before a major operation and another drug, sa\~{n}j\={\i}vini, was
employed to resuscitate the patient after operation or shock.

\subsubsection*{Diagnosis}

It was enjoined that diagnosis be done using all five senses together with 
interrogation.  The diagnosis was based on: 1. cause (nid\={a}na); 
2.premonitory indications (purvar\={u}pa); 3. symptoms (r\={u}pa); 
4. therapeutic tests (upa\'{s}aya); and 5. the natural 
course of development of the disease (sampr\={a}pti). Su\'{s}ruta 
declares that the physician (bhi\d{s}aj), the drug (dravya),  the nurse 
(paric\={a}raka), and the patient (rog\={\i}) are the four pillars on which 
rest the success of the treatment.

Different methods of treatment, based on the diagnosis of the patient, were 
outlined. The drugs were classified into 75 types according to their 
therapeutic effect. For successful treatment, the following ten factors 
were to be kept in mind: 1) the organism (\'{s}ar\={\i}ra); 2) its maintenance 
(v\d{r}tti); 3) cause of disease (hetu); 4) nature of disease (vy\={a}dhi); 5) 
action or treatment (karma); 6) effects or results (k\={a}rya); 
7) time (k\={a}la); 
8) agent or the physician (kart\={a}); 9) the means and instruments (kara\d{n}a); 
and 10) the decision on the line of treatment (vidhi vini\'{s}caya).

Su\'{s}ruta considers the head as the centre of the senses and describes
cranial nerves associated with specific sensory function.  Based on the
derangement of the do\d{s}as, he classifies a total of 1120 diseases.
Caraka, on the other hand, considers the diseases to be innumerable.
The do\d{s}a-type diseases are called {\it nija}, whereas those with an external
basis are called {\it \={a}gantuka}. The microbial origin of disease and the
infective nature of diseases such as fevers, leprosy, smallpox, and
tuberculosis was known. According to Su\'{s}ruta, all forms of leprosy,
some other skin conditions, tuberculosis, opthalmic and epidemic
diseases are born by air and water and may be transmitted from one
person to another.  These diseases are not only due to the derangement
of v\={a}yu, pitta, and kapha, but also of parasitic origin. He adds:
``There are fine organisms that circulate in the blood and are invisible
to the naked eye which give rise to many diseases.''
Parasites were classified into five types:
sahaja (symbiotic parasites), par\={\i}\d{s}aja (derived from faeces),
kaphaja (derived from mucus), \'{s}o\d{n}itaja (derived from the blood
stream), and malaja (derived from the waste products of the body).

Another classification, based on etiological factors, divided disease into 
seven categories: 1) hereditary conditions based on the diseased germ cells 
(\={a}dibala); 2) congenital disease (janmabala); 3) diseases due to the 
disturbance of the humours (do\d{s}abala); 4) injuries and traumas 
(sangh\={a}tabala); 5) sasonal diseases (k\={a}labala); 6) random diseases 
(daivabala); and 7) natural conditions such as aging (svabh\={a}vabala).

The diseases of the head and the nervous system were given in detail. 
Amongst the nervous disorders described are convulsions, apoplectic fits, 
hysteric fits, tetanus, dorsal bending, hemiplegia, total paralysis, facial 
paralysis, lockjaw, stiff neck, paralysis of the tongue, sciatica, St. 
Vitus' dance, paralysis agitans, and fainting. Four kinds of epilepsy was 
described. It was instructed that once the attack is over, the patient 
should not be rebuked and he should be cheered with friendly talk.

Su\'{s}ruta devoted one complete chapter to interpretation of dreams, 
believing that dreams of the patient, together with other omens can be an 
indication to the outcome of the treatment.

\={A}yurveda was also applied to animal welfare. Texts on veterinary
science describe the application of the science to different animals.
Refuges and homes for sick and aged animals and birds were endowed.
An \={a}yurveda for plants and trees was also practised.

To summarize, Indian science was multifaceted, with abstract theories
on the one hand 
and strict protocols for experiments on the other, as in medicine.
Mathematics was used not only in astronomy and music but also in
humanistic subjects such as language.

\section{Cosmology of Greek science}

The Greek tradition, like the Indian, is pluralistic.
Amongst the cosmologies described are: (i) a structure ordered
by a supreme principle (Plato); (ii) cosmos as a
balance of equal but opposed forces (e.g. Anaximander,
Parmenides, and Empedocles); (iii) a cosmos where
war and strife are universal (Homer).

The idea that the universe is a machine under the guidance
of a rational intelligence became prominent at a point.
Plato and Aristotle supported this view, though with different
emphasis. For Plato the Craftsman is transcendent, whereas for
Aristotle it is an immanent force and nature itself is purposeful.
Later philosophers denied the idea that the universe is a product
of design. The atomists and the Epicureans believed that the
world was a product not of design, but necessity out of the
mechanical interactions of atoms. 

There were multiple views about the creation of the universe.
Plato (427-347 BC) appeared to support the view that the universe is one
and created; for Aristotle, it was one and eternal;
for Empedocles, it goes through cycles of creation and destruction.
But these speculations were not associated with specific
numbers as in the great cycles of Indian cosmology.

The early Greeks came into contact with older civilizations and learned
their mathematics and cosmologies. 
Thales of Miletus (born about 624 BC), listed later as the first philosopher,
went to Egypt where he learned  geometry which he
introduced to Greece. He believed that the earth floats 
on water.
His student
Anaximander
believed the earth to be surrounded by a series of spheres made of mist
and surrounded by a big fire.  In a different version of his cosmology he
imagined the earth to be a cylinder floating in space. 

Empedocles believed the cosmos to be egg-shaped and
governed by alternating reigns of love and hate.
He took all matter to be composed of
four elements: earth,
water, fire, air. These four elements arise from the working of the two
properties of hotness (and its contrary coldness) and dryness (and its
contrary wetness) upon an original unqualified or primitive matter. The
possible combinations of these two properties of primitive matter give
rise to the four elements or elemental forms.
In another theory,
Anaximines claimed that everything was made of air.
Earth was some sort of
condensation of air, while fire was some sort of emission form air.
When earth condenses out of air, fire is created in the process.

Pythagoras (560-480 BC) was born on the isle of Samos off the coast of 
Asia Minor. He
travelled widely and studied with priests and healers in several
foreign lands, collecting their  
writings and manuscripts. Between the age of 30 and 40 he
he went into exile from Samos because he did not want to live 
under the rule of Polycrates, the tyrant. He settled down 
in the Greek colony in 
southern Italy called Croton, where he finally established the Pythagorean 
School. 
The Pythagoreans believed in
reincarnation and they were vegetarians.

The Pythagoreans symbolized the elements as geometric forms as part of 
the belief that numbers were the language of physics and psychology. The 
ether element (the sphere of the whole)
is represented by neutrality, spaciousness, and invisibility 
and its qualities were symbolized by the twelve faced dodecahedron. 
The air element is cold, light, quick and it is symbolized by a small blue 
eight faced octahedron. The symbol of the fire element is a very small hot, 
red, active, four faced tetrahedron, and that of water is a larger, moist, 
white 20 faced, white icosahedron. The earth is symbolized by a large, dry, 
yellow, six faced hexahedron. The fire is the smallest and most active, the 
air is slightly larger than fire and very light, the water is slightly 
larger and heavier, and the earth is the largest and most dense. Later, in 
the Timaeus, Plato proposed that the geometric solids constituted the 
elemental shapes of the physical atoms of matter and he based his theory of 
physics on the qualities of their sizes and shapes.

The Greek naturalists called the dynamic vital force the fiery pneuma, 
which is the primordial energy that pervades all phenomena. The expansions 
and contractions of this pneuma produces a space that includes hot 
and cold areas, as well as light and heavy areas of concentration. These 
universal elements manifest as the five archytypal elements ether, air, 
fire, water, earth. The constant ebb and flow of these primordial five 
elements created the interchange of mass and energy which manifested the 
galaxies, solar systems and planets. The flux of these elements produces 
the phenomena of nature we observe as the day and night, the waxing and 
waning of the moon, the passing of the four seasons and the processes of 
biological life itself.

But ordinary matter was constituted of only four elements,
the fifth represented the heavens. 

\subsubsection*{Medicine}
The Pythagoreans believed that 
just as the universe has its central fire, the human body has 
its essence in heat; the heat of the seed and of the uterus are the origin 
of all life; the body attracts to itself the external air on account of its 
desire that heat should be tempered by cold and thus resolves itself in 
respiration.

Empedocles, who proposed the four-element theory, was the founder 
of the Italian school of medicine.
He also noted that the blood and the respiration of air were linked and 
noted that the pores of the skin also breathed. Parmenides 
(450 BC) also wrote works on physiology and psychology. He believed that 
the mental and emotional state of a person was determined by the proportion 
of heat and cold in the human body.

The exchange of the cooling external air and the firey internal heat 
is the cause of both respiration and circulation which stimulates the 
absorption and elimination of liquid and solid nourishment. The synergistic 
combination of the five elements and the pneumatic vital force produces the 
four humoural constituents from air, food and water. These four humours are 
the cold and dry bile, the hot and moist blood, the moist and cold 
phlegm, and the dry and hot choler.

In Greek medicine, 
any excess or deficiency of the four elements and four humours,
disruptions of the three energies, wind (breath), heat (bile) 
and cold (phlegm) produce the state of disease. These factors show that the 
Pythagorean's theory of medicine has much in common with the \={A}yurveda 
which uses the five elements, together with 
the tridh\={a}tu (the three forces) and trido\d{s}a (the three faults).

The most famous person in Greek medicine is Hippocrates, a contemporary
of Plato, known now for the oath that the doctors had to take.
Many texts attributed to Hippocrates shed light upon the Hippocratic method of medicine. None of these
texts may be identified as Hippocrates' own work, however. These works are called the Corpus Hippocraticum
and number upwards of sixty. 
Galen, practicing medicine at Rome in
the latter half of the Second Century AD, was certain, that Hippocrates 
himself wrote
Epidemics I and III.
Scholars have suggested that the texts may have been part of a library
collection, originally from Cos, that was subsequently moved to Alexandria and then added upon, building the
collection of medical texts we have today. While not primary sources, these works were written by
Hippocrates' students and practitioners of his medical theory. 

The Hippocratic physician's 
first option was to suggest a regimen to his patient. 
This treatment consisted of advice regarding what the
patient should eat and drink, and the amount of sleep and exercise he needed; 
it amounted to a cautious ``let
nature take its course'' approach. The purpose of the 
regimen was to void the body of the imbalanced humour
through a diet and exercise program. The physician 
would tailor such a regimen to the time of the year and to
specific patient characteristics, and would wait 
until his patient's condition noticeably improved or worsened. 
The Hippocratic text Aphorisms outlined the order that the treatments were to
follow: ``What drugs will not cure, the knife will; what the knife will not cure, the cautery will; what the cautery
will not cure must be considered incurable.'' 

The choice of drugs to be administered was predicated upon the
perceived humoural imbalance the physician sought to correct. 
The drug of choice was hellebore, popular
because it induced both vomiting and diarrhea, tangible side effects the physician would interpret as the
voiding of the humor. 
Hellebore, an extremely poisonous plant sometimes killed the patient; those
lucky enough to quickly void themselves of its poison managed to survive. 
The physician could justify such
drastic treatment because he believed that the purged material indicated he had successfully balanced the
offending humor and thus treated his patient.

Not surprisingly, this treatment also left the patient in worse health, 
so the physician adopted even more
aggressive methods to cure him. The physician 
resorted to the knife and blood-letting as a solution to the
illness. Venesection was practiced ostensibly because it allowed 
the imbalanced humor a means of directly
escaping the body, thereby restoring the humoral balance and the person's health. The physician would cut
furthest away from the point that hurt, drawing the humours away from that painful spot. 
He would choose
the amount of blood he wanted to spurt out by placing a cup over the bleeding incision; once the cup had filled,
the physician was satisfied that he had completed treatment.

The physician's last-ditch effort was cauterization. 
Assuming that the patient had
not yet died from either the disease or the previous treatments, 
cauterization involved burning the skin in one
last attempt to ``consume" the excess humor. 
Then the physician would allow the resulting wound to ulcerate,
which he would then irritate with caustic drugs, like mustard-seed paste, 
to allow the humours to slowly drain
out of the body. 
The area chosen for cauterization was to be located as far as possible 
from the actual
wound, again to draw the excess humours from the 
painful spot.

In common with other intellectuals in the Greek city-states, Hippocratics were 
interested in
far-away places and peoples, in epidemic diseases and plagues, in the origins of
man and embryology, and in dietetics. 
Hippocratics were quick to criticize causes and remedies that they considered
irrational. 
The writer of ``Diseases of Young Girls'' censures women who follow 
commands from Artemis' priests
to dedicate costly garments to the goddess in the effort to cure 
madness in the premenarchic young girl. 
The author of ``Sacred Disease''
criticizes ``witch-doctors, faith-healers, quacks and charlatans," whose etiology for epilepsy and
sudden seizures invokes attacks from the gods and whose therapies consist of purifications,
incantations, prohibition of baths, lying on goat-skins and eating goats' flesh 
({\it Sacred Disease 1-2}).

The etiology for the disease in such examples is taken to be 
the blockage of inner vessels by a bodily humour.
Treatment was to require the evacuation of the
noxious fluid from vital areas of the body: the epileptic is to take a 
medicine to move excess
phlegm gradually from his head so that its sudden descent into his body doesn't overwhelm his
senses, and the young girl is to sleep with a man as soon as possible to remove the impediment at
the mouth of her uterus, while pregnancy will bring her long-lasting cure by opening up her body so
that her excess fluids can move about freely. 

Although one may criticize these methods, there was a basis
of process that was invoked to explain a disease. This then
qualified the system as being scientific. The Hippocratic system was to remain
extremely influential in the West until the advent of modern medicine.

\subsubsection*{Elements and the solar system}
We have already mentioned Plato's
mathematical
construction of the elements (earth, fire, air, and water), in which the cube, 
tetrahedron, octahedron, and
icosahedron are given as the shapes of the atoms of earth, fire, air, 
and water. The fifth Platonic solid, the
dodecahedron, is Plato's model for the whole universe. 

Plato's beliefs as regards the universe were that the stars, planets, the
sun and the
moon move round the earth in
crystalline spheres. The sphere of the moon was closest to the earth, then the 
sphere of the sun, then Mercury,
Venus, Mars, Jupiter, Saturn and furthest away was the sphere of the stars. 
He believed that the moon shines by reflected sunlight.

Elements had a natural tendency to separate in space; fire moved
outwards, away from the earth, and earth moved inwards, with air and
water being intermediate. Thus, each of these elements occupied a
unique place in the heavens (earth elements were heavy and, therefore,
low; fire elements were light and located up high).

There were only seven objects visible to the ancients, the sun and the
moon, plus the five planets, Mercury, Venus, Mars, Jupiter and Saturn.
It was obvious that the planets were not on the celestial sphere since
the moon clearly passes in front of the sun and planets, plus Mercury
and Venus can be seen to transit the sun. 

Slightly later, Aristarchus (270 B.C.) proposed an alternative model of
the Solar System placing the sun at the center with the earth and the
planets in circular orbit around it. The moon orbits around the earth.
This model became known as the heliocentric theory
Aristarchus' model was ruled out by the philosophers at the
time for the reason that there was no feeling on earth of its motion.
It is interesting that although Aristarchus was the most prominent
astronomer of his time, he assumed a figure of two degrees for
the angular diameter of the moon, when the correct value is about
one-half of a degree. Since the correct size is very easily 
ascertained, the criticism has been made that the Greeks did not
have a tradition of observational astronomy.

\subsubsection*{Ptolemaic system}

Ptolemy (200 A.D.) 
was the author
of the Almagest, a treatise on the celestial sphere and the motion
of the planets. The book is divided into 13
books, each of which deals with certain astronomical concepts
pertaining to stars and to objects in the solar system. It was, no
doubt, the encyclopedic nature of the work that made the Almagest so
useful to later astronomers and that gave the views contained in it so
profound an influence. In essence, it is a synthesis of the results
obtained by Greek astronomy; it is also the major source of knowledge
about the work of Hipparchus.

In the first book of the Almagest, Ptolemy describes his geocentric
system and gives various arguments to prove that, in its position at
the center of the universe, the earth must be immovable. Not least, he
showed that if the earth moved, as some earlier philosophers had
suggested, then certain phenomena should in consequence be observed. In
particular, Ptolemy argued that since all bodies fall to the center of
the universe, the earth must be fixed there at the center, otherwise
falling objects would not be seen to drop toward the center of the
earth. Again, if the earth rotated once every 24 hours, a body thrown
vertically upward should not fall back to the same place, as it was
seen to do. Ptolemy was able to demonstrate, however, that no contrary
observations had ever been obtained.

Ptolemy accepted the following order for celestial objects in the solar
system: earth (center), Moon, Mercury, Venus, Sun, Mars, Jupiter, and
Saturn. In particular, the sun appears to describe a yearly circular
path called the ecliptic over the background of stars. However, when
the detailed observations of the planets in the skies is examined, the
planets undergo motion which is impossible to explain in the geocentric
model, a backward track for the outer planets. This behavior is called
retrograde motion.

He realized, as had Hipparchus, that the inequalities in the motions of
these heavenly bodies necessitated either a system of deferents and
epicycles or one of movable eccentrics (both systems devised by
Apollonius of Perga, the Greek geometer of the 3rd century BC) in order
to account for their movements in terms of uniform circular motion.
In the Ptolemaic system, deferents were large circles centered on the
earth, and epicycles were small circles whose centers moved around the
circumferences of the deferents. The sun, moon, and planets moved
around the circumference of their own epicycles. In the movable
eccentric, there was one circle; this was centered on a point displaced
from the earth, with the planet moving around the circumference. These
were mathematically equivalent schemes.

Even with these, all observed planetary phenomena still could not be
fully taken into account. Ptolemy now
supposed that the
earth was located a short distance from the center of the deferent for
each planet and that the center of the planet's deferent and the
epicycle described uniform circular motion around what he called the
equant, which was an imaginary point that he placed on the diameter of
the deferent but at a position opposite to that of the earth from the
center of the deferent (i.e., the center of the deferent was between
the earth and the equant). He further supposed that the distance from
the earth to the center of the deferent was equal to the distance from
the center of the deferent to the equant. With this hypothesis, Ptolemy
could better account for many observed planetary phenomena.

\subsubsection*{Crystalline spheres}

Although Ptolemy realized that the planets were much closer to the
earth than the ``fixed" stars, he believed in the physical
existence of Plato's crystalline spheres, to which the heavenly bodies were
said to be attached. Outside the sphere of the fixed stars, Ptolemy
proposed other spheres, ending with the primum mobile (``prime mover"),
which provided the motive power for the remaining spheres that
constituted his conception of the universe. 

This model, while complicated, was a complete description of the Solar
System that explained, and predicted, the apparent motions of all the
planets. 
This, however was not universally accepted. The most notable detractor
was Democritus who postulated the existence of indestructible atoms
(from the Greek a-tome: that which cannot be cut) of an infinite
variety of shapes and sizes. He imagined an infinite universe
containing an infinite number of such atoms, in between the atoms there
is an absolute void.

\subsubsection*{Aristotle's cosmology}

Aristotle's cosmological work {\it On The Heavens} presents 
the mainstream view of Greek cosmology.
Aristotle believed in just four elements: earth,
water, air and fire. These elements naturally move up or
down, fire being the lightest and earth the heaviest. A composite
object will have the features of the element which dominates.

The idea that all bodies, by their very nature, have a natural way of
moving is central to Aristotelian cosmology.  Movement is not, he
states, the result of the influence of one body on another
Some bodies naturally move in straight lines, others naturally stay
put. But there is yet another natural movement: the circular motion.
Since to each motion there must correspond a substance, there ought to
be some things that naturally move in circles. Aristotle then states
that such things are the heavenly bodies which are made of a more
exalted and perfect substance than all earthly objects.

Since the stars and planets are made of this exalted substance and then
move in circles, it is also natural, according to Aristotle, for these
objects to be spheres. The cosmos is then made of a central earth
(which he accepted as spherical) surrounded by the moon, sun and stars
all moving in circles around it. This conglomerate he called ``the
world''. Although celestial bodies are perfect,
they must circle the imperfect earth. The initial motion of these
spheres was caused by the action of a ``prime mover'' which acts
on the outermost sphere of the fixed stars; the motion then trickles
down to the other spheres through a dragging force.

Aristotle also addresses the question whether this world is unique or
not; he argues that it is unique. The argument goes as follows: earth
(the substance) moves naturally to the center, if the world is not
unique there ought to be at least two centers, but then, how can earth
know to which of the two centers to go? But since ``earthy'' objects
have no trouble deciding how to move, he concludes that there can only
be one center (the earth) circled endlessly by all heavenly bodies. 

It is interesting to note that Aristotle asserts that the world did not
come into being at one point, but that it has existed, unchanged, for
all eternity (it had to be that way since it was ``perfect''); the
universe is in a kind of ``steady state scenario''. Still, since he
believed that the sphere was the most perfect of the geometrical
shapes, the universe did have a center (the earth) and its ``material''
part had an edge, which was ``gradual'' starting in the lunar and
ending in the fixed star sphere. Beyond the sphere of the stars the
universe continued into the spiritual realm where material things
cannot be. 

On the specific description of the heavens, Aristotle created a complex
system containing 55 spheres.
One of the fundamental propositions of Aristotelian
philosophy is that there is no effect without a cause.  Applied to
moving bodies, this proposition dictates that there is no motion
without a force. Speed, then is proportional to force and inversely
proportional to resistance.

Qualitatively this implies that a body will traverse a thinner medium
in a shorter time than a thicker medium (of the same length): things
will go faster through air than through water. A natural (though
erroneous) conclusion is that there could be no vacuum in Nature, for
if the resistance became vanishingly small, a tiny force would produce
a very large ``motion''; in the limit where there is no resistance any
force on any body would produce an infinite speed. This conclusion put
him in direct contradiction with the ideas of the atomists such as
Democritus. Aristotle concluded the
atomists were wrong, stating that matter is in fact continuous and
infinitely divisible.

For falling bodies, the force is the weight pulling down a body and the
resistance is that of the medium (air, water, etc.). Aristotle noted
that a falling object gains speed, which he then attributed to a gain
in weight. If weight determines the speed of fall, then when two
different weights are dropped from a high place the heavier will fall
faster and the lighter slower, in proportion to the two weights. A ten
pound weight would reach the earth by the time a one-pound weight had
fallen one-tenth as far.

\subsubsection*{Zeno's paradoxes}
An excellent example of the
Greek application of logic in understanding motion is captured by
the paradoxes of Zeno (fifth century BC). 
These paradoxes, which are variations
on a single theme, emerge because of the consideration of space and
time as discrete.

\begin{enumerate}

\item
{\it Dichotomy paradox:} Before a moving object can travel a certain distance, 
it must travel half that distance.
     Before it can travel half the distance it must travel one-fourth 
the distance, 
and so on. This sequence goes on forever,
     therefore, the original distance cannot be traveled, 
and motion is impossible. 
  \item
{\it Achilles and the tortoise paradox:} Achilles gives the tortoise a head 
start in a race. Before he can
     overtake the tortoise, he must run to the place where the tortoise began
but the tortoise has moved on to
     some other point. From there, before he can overtake the tortoise, 
he must run to the place where the
     tortoise had moved to. This goes on forever, and Achilles can never pass the tortoise. 
  \item
{\it Arrow paradox:} If you look at an arrow in flight, at an instant in time, it appears the same as a motionless
     arrow. Then how does it move?

\end{enumerate}
\subsubsection*{Archimedes and physics }

Archimedes (3rd century BC)
contributed many new results to mathematics, including
successfully computing areas and volumes of two and three dimensional
figures and a geometrical argument for an approximation of
$\pi$.
His major contributions to physics are his 
principle of buoyancy, and his analysis of the lever. He also invented many
ingenious technological devices, many for war, but also the Archimedean
screw, a pumping device for irrigation systems.
 
According to
the Roman historian Vitruvius, 
Hiero, after gaining the royal power in Syracuse, resolved, as a
consequence of his successful exploits, to place in a certain temple a
golden crown which he had vowed to the immortal gods. He contracted for its
making at a fixed price and weighed out a precise amount of gold to the
contractor. At the appointed time the latter delivered to the king's
satisfaction an exquisitely finished piece of handiwork, and it appeared
that in weight the crown corresponded precisely to what the gold had weighed.
But afterwards a charge was made that gold had been abstracted and an
equivalent weight of silver had been added in the manufacture of the crown.
Hiero asked his friend 
Archimedes to investigate the matter. While in a tub for his bath,
Archimedes discovered the principle of buoyancy (hydrostatics or the
Archimedes Law), according to which the water displaced equals
the mass of the body.
Without a moments delay and transported with joy, he
jumped out of the tub and rushed home naked, crying in a loud voice 
``Eureka, Eureka."

Archimedes' Principle states that the buoyancy support force is exactly
equal to the weight of the water displaced by the immersed object, that is, it is
equal to the weight of a volume of water equal to the volume of the 
object.
 
Although leverage has been used to move heavy objects since
prehistoric times, it appears that Archimedes was the first person to
appreciate just how much weight could be shifted by one person using
appropriate leverage.
It is believed that he also invented a screw for pumping water upwards.

\section{Comparing Indian and Greek sciences}

We first note that when we speak of a formal science, more than
a list of of general ideas it is 
the elaboration of a system.
Using this definition, medicine is the best examples of a full-fledged science
from the ancient world.
The Greek and Indian medical sciences have some points of
commonality, but they also have distinct differences.
The Greek system emphasizes the mechanical aspects of humours,
whereas the Indian system considers connections with the
mind as well. The Indian system has many recursive levels,
it has a psychosomatic concept of disease, and
it is a more comprehensive medical system.
My personal judgment is that \={A}yurveda is superior to
the Hippocratic system which is why it is still practiced in
India and the West. 

In other fields, the level of knowledge varied.
Considering the work of Archimedes, it is clear that
Greek knowledge of hydrostatics was more advanced.
On the other hand, Indian inner sciences
(psychology) were more sophisticated and the Indians had 
greater success in grammar
and in the conception of the physical universe.

The differences between the two traditions arose from
their respective cosmologies, which is why they
did not attempt to learn from each other. 
The Indian cosmos was infinite with a postulated connection
between the inner and the outer. The Greek cosmos was a finite
system.

\subsubsection*{Greek priority}
This view flies against great
mass of evidence.
For example, the references to the three humours in the Atharvaveda
are prior to Plato by any reckoning.
Likewise, the Baudh\={a}yana \'{S}ulbas\={u}tra is prior to 
Greek geometry. 
Indian astronomy is also earlier as evidenced by 
its description in the \'{S}atapatha Br\={a}hma\d{n}a and
the Ved\={a}\.{n}ga Jyoti\d{s}a.

For some time the \'{S}ulbas\={u}tras themselves were considered
to be very late, after Euclid, in spite of the powerful 
linguistic arguments against this view.
But when it was discovered the same geometry was present
in the \'{S}atapatha Br\={a}hma\d{n}a, which is dated prior to
600 BC in the most conservative chronologies, 
the circular logic of such
chronology became evident.$^{20}$
Similar misguided logic
was used to assign the date
of 500 BC to  
the Ved\={a}\.{n}ga Jyoti\d{s}a, which has an inner date of
roughly 1300 BC, to make its
astronomy dependent on Greek and Babylonian sources.

\subsubsection*{Common prior origin}

This theory comes in two main forms. 
In the first one, which is a variant of the standard Eurocentric
view, it is conceded that the Greek culture was 
indebted to the Mesopotamians and the Egyptians, and these
taken together form the basis of Western science.

The second version includes Indian evidence.
The \'{S}ulbas\={u}tra geometry is different from
the Mesopotamian one in that it is constructive and not
purely algebraic and it shares features with Greek geometry.
Therefore, if there is
common origin that must be in the shared Indo-European
heritage.
In other words, this geometric knowledge must have been a
part of the itinerant Indo-European tribes as they moved
through Central Asia, ending up later as the Vedic Indians and
the Greeks.$^{21}$

But the Greek tradition doesn't know 
archaic aspects of Vedic science which
are more central to the knowledge system than 
geometrical knowledge.
Therefore the idea of common prior origin must be
discounted.

\subsubsection*{Indian priority}

There are elements of the Greek system which were clearly
borrowed from outside. 
Thales and Pythagoras travelled out of the country and
introduced new ideas. The Pythagorean
ideas on five elements, vegetarianism and reincarnation
appear to have an Indian origin.
The Pythagoras theorem of geometry may have had an
Indian origin since it is described in the \'{S}ulba S\={u}tras.
However, such broad concepts do not make a system. 
It is the amplification of these ideas which reflects
the national genius.

The history of Greek science is well understood. It 
is clear that while it may have been inspired initially by foreign
impulses, it adopted a unique trajectory and there is
no evidence that it was in any way dependent on Indian science later.
In astronomy, for example, there is no evidence of the use
of Indian methods.
Greek medicine is likewise different from Indian medicine in
crucial details.

\subsubsection*{Independent development}

The physics, philosophy, and
medicine of the Greeks was undoubtedly a result
of the characteristic elaboration within Greece.
Even the idea of five elements did not become
the mainstream view and the Hippocratics and Aristotle
spoke only of
four elements.

The Indian system of five elements is connected to
the five-fold recursive division of reality in
S\={a}\d{m}khya. On the other hand, the 
emergence of a fifth element in Greek thought was a gradual process.
In the common conception of the universe 
shared by most religious and
philosophical thinkers in the centuries before Plato, the cosmos, a sphere bounded by the sky, contains the
conflicting "opposites" (the hot/the cold, the wet/the dry) 
which became (via Empedocles) the four
root substances earth, water, air and fire.

The astronomies of India and Greece are quite different.
The elaboration of logic is also distinctively different.
In view of all of this,
the developments of Greek and Indian
science must be considered independent.$^{22}$

\section{Zodiacal signs and Greek astronomy in India}

The argument for a Greek origin of the later Indian astronomy
was articulated most vehemently by W.D. Whitney in the
closing decades of the 19th century.
He suggested that the transmission of Greek ideas
took place sometime before Ptolemy because the
Indian methods do not mention the techniques
introduced by him.
He also suggested that this transmission must have occurred
in consequence of the ``lively commerce between
Alexandria as the port and mart of Europe and the
western coast of India [during the first centuries
of our era].''$^{23}$
He was certain that India could not have been the place
of the origin of Indian astronomy because he believed that
Indian history did not go further back than 2000 BC.

Whitney's theory
hinges around
the use of the words {\it lipt\={a}}, {\it hor\={a}},
and {\it kendra} for minute, hour, and the mean anomaly,
which are supposed to be of Greek origin.
Furthermore, he believed that there is scant mention
of the planets in early texts, the names of
the days after the planets, and the division of the
circle into signs, degrees, minutes, and seconds
is fashioned after Greek usage.

E. Burgess countered Whitney with arguments related
to  the lunar and solar divisions of the zodiac, the epicycle
theories, astrology, and the names of the five planets.$^{24}$
I summarize below these arguments strengthened with the new
insights that have been obtained by recent research:$^{25}$

\begin{description}

\item
{\it 1. Lunar zodiac.}
The demonstration that the \d{R}gveda itself contains the
list of the 27 nak\d{s}atras in terms of their
presiding deities$^{26}$ establishes that the
lunar zodiac has a greater antiquity in India as compared
to the Chinese and the Arabs.
Furthermore, the beginnings of the Indian tradition
are now believed to be several thousand years earlier than
thought by Whitney.

\item
{\it 2. Solar zodiac.}
I have recently shown$^{27}$ that the solar zodiac arose in
India since the names and symbols of the signs are intimately
related to the deities of the corresponding nak\d{s}atras,
whereas
they appear out of context in Babylonia and Greece.
Burgess anticipated this when he stated that this twelve-part
``division was known to the Hindus centuries before any trace
can be found in existence among any other people.''

\item
{\it 3. The theory of epicycles.}
I concur with Burgess's judgment that ``the difference of
this theory in the Greek and Hindu systems of astronomy
precludes the diea that one of these people derived more than
a hint respecting it from the other. And so far as this point
alone is concerned, we have as much reason to suppose the Greeks
to have been the borrowers as the contrary; but other 
considerations seem to favor the supposition that the Hindus 
were the original inventors of this thory.''

\item
{\it 4. Coincidence in the systems of astrology.}
According to Burgess, ``The coincidence that exist between
the Hindu and Greek systems are too remarkable to admit of
the supposition of an independent origin for them.
But the honor of original invention, such as it is, lies,
I think, between the Hindus and the Chaldeans. The evidence
of priority of invention and culture seems, on the whole, to
be in favor of the former; the existence of three or
four Arabic and Greek terms in the Hindu system being
accounted for on the supposition that they were introduced
at a comparatively recent period. In reference to the
word hor\={a} it may not be inappropriate to introduce the
testimony of Herodotus: `The sun-dial and the gnomon,
with the division of the day into twelve parts, were
received by the Greeks from the Babylonians.' There
is abundant testimony to the fact that the division of
the day into twenty-four hours existed in the East, if
not actually in India, before it did in Greece. In reference
to the so-called Greek words found in Hindu astronomical
treatises, I would remark that we may with entire
propriety refer them to the numerous class of words common to
the Greek and Sanskrit languages, which either come to both
from a common source, or passed from the Sanskrit to the
Greek at a period of high antiquity; for no one maintains,
so far as I am aware, that the Greek is the parent of
Sanskrit.''

An examination of the Vedic texts completely settles this question.$^{28}$
We have overwhelming evidence supporting the view that
the twelve part division of the circle goes back to
the \d{R}gvedic period itself.
The Indians derive the word hor\={a} from 
ahor\={a}tra (day-night) which is attested in the Vedic texts.

\item
{\it 5. Names of planets.} Burgess argued
that the application of the names of the planets to
the days of the week was unknown to the Greeks and
not adopted by the Romans until a late period.
He concurred with H.H. Wilson, who said: ``It is 
commonly ascribed to the Egyptians and Babylonians,
but upon no very sufficient authority, and the Hindus
appear to have at least as good a title to the
invention as any other people.''

The planets are actually mentioned in the Vedic texts.$^{29}$

\end{description}

Burgess emphasized that the Arabs were thoroughly 
imbued with the knowledge of the Hindu astronomy before
they became acquainted with that of the Greeks.
This is established by the Arab translations of
Ptolemy's Syntaxis (Almagest). In the Latin translations of this
book, the ascending node is called {\it nodus capitis,}
``the node of the head,'' and the descending node 
{\it nodus caudae,} ``the node of the tail (ketu),''
which are pure Hindu appellations.

Burgess concluded by speaking of the detailed methods:
``In the amount of the annual precession of the 
equinoxes, the relative size of the sun and moon as
compated with the earth, the greatest equation of the
centre for the sun -- the Hindus are more nearly correct
than the Greeks, and in regard to the times of the
revolutions of the planets they are very nearly correct in
four itmes, and Ptolemy in six. There has evidently been
very little astronomical borrowing between the Hindus and
the Greeks. And in regard to points that prove a communication
from one people to the other, I am inclined to think that
the course of derivation was from east to west rather than
from west to east.''

The opinions of Burgess are strengthened by our improved knowledge
of the astronomy of the Ved\={a}\.{n}ga Jyoti\d{s}a and
the discovery of the astronomy of the Vedic fire altars.$^{30}$
Neither should we forget that the cosmological basis of the
Indian sciences was much more subtle and comprehensive than
that of Greek science. The achievements of Indian sciences
in the fields of mathematics, grammar, and the 
comprehensiveness of its medical science also show that
the Indian sciences were more advanced that the corresponding
Greek sciences during the glory days of Greek civilization.

The thesis that Indian siddh\={a}ntas were based on
Greek and Babylonian measurements has been fully refuted by
Billard and van der Waerden and so it will not be discussed here.$^{31}$
It is noteworthy that the framework of the siddh\={a}ntas was very
different from that of Greek models.
For example, \={A}ryabha\d{t}a's astronomy considers the
earth to rotate on its axis and the planets to go around the sun,
which, in turn, is taken to go around the earth.
This model was more advanced than contemporaneous models in the
ancient world.$^{32}$

\section{Conclusions}

We have
indirect evidence
that there was a period of interaction between Indian and Greek science
before the flowering of Greek philosophy.
This was the time when the Greeks borrowed from many sources that
apparently included the Indian.
But the importance of this borrowing must not be exaggerated
since it merely consisted of general ideas such as
five elements of reality or the 360-part division of the circle.

The Greeks were fully aware of the Asiatic origin of their ideas.
For example, Strabo informs us$^{33}$ that music ``from the triple
point of view of melody, rhythm and instruments'' came to them
originally from Thrace and Asia. Further, ``the poets, who make 
the whole of Asia, including India, the land or sacred territory
of Dionysos, claim that the origin of music is almost entirely
Asiatic.'' Study of music and mathematics go hand in hand, showing
the Greek awareness of the Eastern connection with their own
traditions.

Subsequent to that, Greek and Indian sciences appear to have
developed independently although after the time of Alexander they
were consciously aware of each other.
The focus and style of the two sciences was different owing
to their different cosmologies. Perhaps the
only science which worked more or less the same way in
the two civilizations was medicine.
But even here there were important differences.

The reason why the two sciences went their own way in spite
of the knowledge of the other was because their worldviews
 were different.
Indian astronomers, for example, never paid any attention to
Ptolemy's model as its crystalline sphere basis looked
very primitive compared to the vastness of their own 
conception.
On the other hand, the use of enormous time scales of Indian
astronomy must have appeared unnecessary to the Greeks.

There was no specific technology that arose at this early date that might
have caused India or Greece to question its own system and become receptive
to new ways of doing things. The power and influence of the Greek
ideas arose out of the narrative and style of its philosophers and
scientists.
Its influence was primarily through its literature and philosophy.
Likewise, Indian influence all over Asia was through its philosophy,
medicine,
and the arts and its narrative texts.
Science could take its next steps only after the development of
new technology which opened up new worlds of the small and the large.

\section*{Notes and References}
\begin{enumerate}

\item S. Kak, ``Indic ideas in the Graeco-Roman world.''
{\it Indian Historical Review}, 1999.

\item
S. Kak, Akhenaten, S\={u}rya, and the \d{R}gveda. LSU, 2003.

\item
Arrian, {\it The Campaigns of Alexander.} Viking, New York, 1976.

\item See, for example, G.J. Joseph, {\it The Crest of the Peacock:
Non-European Roots of Mathematics.} Princeton University Press, 2000.

\item 
Joseph, {\it op cit.}

\item A. Seidenberg, The origin of mathematics.
{\it Archive for History of Exact Sciences.} 18: 301-342, 1978.

\item S. Kak, {\it The Astronomical Code of the \d{R}gveda.}
New Delhi: Munshiram Manoharlal, 2000.

\item J. Filliozat, ``The expansion of Indian medicine abroad.''
In Lokesh Chandra (ed.) {\it India's Contributions to World Thought
and Culture.} Madras: Vivekananda Memorial Committee. 67-70, 1970.

\item W.D. Whitney, Notes at the end of the S\={u}rya Siddh\={a}nta,
translated by E. Burgess.

\item F. Staal, ``Greek and Vedic geometry,'' {\it Journal of Indian 
Philosophy,} 27: 105-127, 1999.
 
\item
D.P. Singhal, {\it India and World Civilization.} Michigan University
Press, 1969.
This book does not argue for priority of Indian science; rather, it
presents much evidence for Indian influence in a variety of fields
on world civilization. It also summarizes the evidence for possible
early links between India and the New World.

\item
E. Burgess (tr.), {\it S\={u}rya Siddh\={a}nta.}

\item
P.C. Sengupta, {\it Indian Chronology.}
University of Calcutta, 1947;
S. Kak, {\it Astronomical Code..}, {\it op cit.}

\item
S. Kak, ``The evolution of writing in India.'' {\it Indian Journal of History
of Science}, 28: 375-388, 1994.\\
That the
ten most common symbols of Indus are structurally similar
to the ten most common symbols of Br\={a}hm\={\i} indicates a genetic 
relationship.  

\item Bh\={a}gavata Pu. 9.3.28-32.

\item 
S. Venkatesananda (ed.), {\it V\={a}si\d{s}\d{t}ha's Yoga.}
State University of New York Press, Albany, 1993. For the Sankrit text, see
{\it Yoga V\={a}si\d{s}\d{t}ha}, 1981. Munshiram Manoharlal, Delhi.

\item B. van Nooten, ``Binary numbers in Indian antiquity'',
{\it Journal of Indian Philosophy}, 21, 31-50, 1993; reprinted in
T.R. N. Rao and Kak, S. (eds.), {\it Computing Science in Ancient India.}
Munshiram Manoharlal, New Delhi, 2000.

\item
See the articles in T.R.N. Rao and S. Kak (eds.),
{\it Computing Science in Ancient India.}
Munshiram Manoharlal, New Delhi, 2000.

\item
S. Kak, {\it The Astronomical Code..}
{\it op cit.}

\item
Seidenberg, {\it op cit.}
S. Kak, ``Babylonian and Indian astronomy: early connections.''
Los Alamos Archive, January 2003.

\item
Staal, {\it op cit.}

\item
Burgess, {\it op cit.}

\item
Whitney, {\it op cit.}

\item
Burgess, {\it op cit.}

\item
N. Achar,  ``On the Vedic origin of the ancient
mathematical astronomy of India.''
{\it Journal of Studies on Ancient India},
1: 95-108, 1998.
Kak, ``Babylonian and Indian astronomy: early connections.''
{\it op cit.}

\item
Achar, {\it op cit.}

\item
Kak, ``Babylonian and Indian astronomy,'' {\it op cit.}

\item
Kak, {\it op cit.}

\item
S. Kak,
``Knowledge of the planets in the third millennium BC," 
{\em Quarterly Journal of the Royal Astronomical Society,} 37, 1996, pp. 709-715.

\item This work may be found collectively in
S. Kak, ``Birth and early development of Indian astronomy,''
In {\it Astronomy Across Cultures: The History of 
Non-Western Astronomy.} H. Selin (ed.). Kluwer Academic,
Boston, 2000, pp. 303-340;\\ 
S. Kak, {\em The Astronomical Code of the \d{R}gveda}. 
Munshiram Manoharlal, Delhi, 2000.

\item R. Billard, {\it L'astronomie Indienne.}
Paris: Publications de l'ecole francaise d'extreme-orient, 1971.
See also 
B.L. van der Waerden, ``Two treatises on Indian
astronomy.'' {\it Journal for History of Astronomy}
11: 50-58, 1980, which concludes that Billard's critics
are wrong and that the siddh\={a}ntas were
based on measurements made within India.

\item
H. Thurston, {\it Early Astronomy.}
New York: Springer-Verlag, 1994, page 188.
Note that Aristarchus' heliocentric model was
only a speculative suggestion and not part of
a fully elaborated sytem.

\item Strabo, {\it Geography,} 10,11,17.

\end{enumerate}

\end{document}